\documentclass[superscriptaddress,twocolumn,amsmath,amssymb]{revtex4-1}
\pdfoutput=1
\usepackage{graphicx}
\usepackage{eurosym}
\usepackage{amsmath}
\usepackage{amssymb}
\usepackage{dcolumn}
\usepackage{bm}
\usepackage{subfigure}
\usepackage[final]{pdfpages}
\usepackage{soul}
\usepackage{algcompatible}
\usepackage{newfloat}

\DeclareFloatingEnvironment[
    fileext=loa,
    listname=List of Algorithms,
    name=ALGORITHM,
    placement=tbhp,
]{algorithm}

\def\bra#1{\mathinner{\langle{#1}|}}
\def\ket#1{\mathinner{|{#1}\rangle}}
\newcommand{\braket}[2]{\langle #1|#2\rangle}

\DeclareMathOperator{\Tr}{Tr}

\newcommand{\ignore}[1]{}

\newcommand{\be}{\begin{equaArXivtion}}
\newcommand{\ee}{\end{equation}}
\newcommand{\ba}{\begin{eqnarray}}
\newcommand{\ea}{\end{eqnarray}}

\begin{document}

\title{Compact wavefunctions from compressed imaginary time evolution}

\author{Jarrod R. McClean} 
\affiliation{Department of Chemistry and Chemical Biology, Harvard University, Cambridge MA, 02138}
\author{Al\'an Aspuru-Guzik}
\affiliation{Department of Chemistry and Chemical Biology, Harvard University, Cambridge MA, 02138}

\begin{abstract} 
Simulation of quantum systems promises to deliver physical and chemical predictions for the frontiers of technology.  Unfortunately, the exact representation of these systems is plagued by the exponential growth of dimension with the number of particles, or colloquially, the curse of dimensionality.  The success of approximation methods has hinged on the relative simplicity of physical systems with respect to the exponentially complex worst case.  Exploiting this relative simplicity has required detailed knowledge of the physical system under study.  In this work, we introduce a general and efficient black box method for many-body quantum systems that utilizes technology from compressed sensing to find the most compact wavefunction possible without detailed knowledge of the system. It is a Multicomponent Adaptive Greedy Iterative Compression (MAGIC) scheme. No knowledge is assumed in the structure of the problem other than correct particle statistics.  This method can be applied to many quantum systems such as spins, qubits, oscillators, or electronic systems.  As an application, we use this technique to compute ground state electronic wavefunctions of hydrogen fluoride and recover 98\% of the basis set correlation energy or equivalently 99.996\% of the total energy with $50$ configurations out of a possible $10^7$.  Building from this compactness, we introduce the idea of nuclear union configuration interaction for improving the description of reaction coordinates and use it to study the dissociation of hydrogen fluoride and the helium dimer.
\end{abstract}

\maketitle

\section{Introduction}
The prediction of chemical, physical, and material properties from first principles has long been the goal of computational scientists.  The Schr\"odinger equation contains the required information for this task, however its exact solution remains intractable for all but the smallest systems, due to the exponentially growing space in which the solutions exist.  To make progress in prediction, many approximate schemes have been developed over the years that treat the problem in some small part of this exponential space.  Some of the more popular methods in both chemistry and physics include Hartree-Fock, approximate density functional theory, valence bond theory, perturbation theory, coupled cluster methods, multi-configurational methods, and more recently density matrix renormalization group~\cite{Slater:1951,Baerends:1973,Parr:1989,Hunt:1972,Goddard:1973,
Moller:1934,Bartlett:1981,Helgaker:2002,White:1999,Chan:2011}.

These methods have been successful in a wide array of problems due largely to the intricate physics they compactly encode.  For example, methods which are essentially exact and scale only polynomial with the size of the system have been developed for one-dimensional gapped quantum systems~\cite{Landau:2013}.  However such structure is not always easy to identify or even present as the size and complexity of the systems grow.  For example, some biologically important transition metal compounds as well as metal clusters lack obvious structure, and remain intractable with current methods.

The field of compressed sensing exploits a general type of structure, namely simplicity or sparsity, which has been empirically observed and is adaptive to the problem at hand.  Recent developments in compressed sensing have revived the notion that Occam's razor is at work in physical systems.  That is, the simplest feasible solution is often the correct one.  Compressed sensing techniques have had success in quantum simulation in the context of localized wavefunctions~\cite{Oz:2013} and vibrational dynamics of quantum systems~\cite{Wu:2003,Chen:2006}, but little has been done to exploit the possibilities for many-body eigenstates, which are critically important in the analysis and study of physical systems.  

In this work, we concisely describe a new methodology for finding compact ground state eigenfunctions for quantum systems.  It is a Multicomponent Adaptive Greedy Iterative Compression (MAGIC) scheme.  This method is general in that it is not restricted to a specific ansatz or type of quantum system.  It operates by expanding the wavefunction with imaginary time evolution, while greedily compressing it with orthogonal matching pursuit~\cite{Tropp:2007}.  As an application, we choose the simplest possible ansatze for quantum chemistry, sums of non-orthogonal determinants, and demonstrate that extremely accurate solutions are possible with very compact wavefunctions. This non-orthogonal MAGIC scheme we refer to as NOMAGIC, and apply it electronic wavefunctions in quantum chemistry.

\section{Compressed imaginary time evolution}
Beginning with general quantum systems, an $N$-particle eigenfunction of a quantum Hamiltonian $H$, $\ket{\Psi}$, may be approximated by a trial function $\ket{\tilde \Psi}$ that is the sum of many-particle component functions $\ket{\Phi^i}$, such that
\begin{align}
 \ket{\tilde \Psi} = \sum_i^{N_c} c_i \ket{\Phi^i}
\end{align}
where $N_c$ is the total number of configurations in the sum and no relation need be assumed between $\ket{\Phi^i}$ and $\ket{\Phi^j}$ for $i \neq j$.  A simple example of such a component function for a general quantum system is the tensor product of $N$ single particle functions $\ket{\phi^i_j}$
\begin{align}
 \ket{\Phi^i} = \ket{\phi_0^i}\ket{\phi_1^i}...\ket{\phi_{N-1}^i}
\end{align}
and we will consider its anti-symmetric counterpart in applications to electronic systems.  In this work, we adopt a state to be simple, sparse, or compact in the total space if the number of configurations $N_c$ needed to represent a state to a desired precision is much less than the total dimension of the Hilbert space.

One method for determining $\ket{\tilde \Psi}$ is a direct variational approach based on the particular parametrization of $\ket{\Phi^i}$ and choice of $N_c$.  This approach can plagued by issues related to the choice of initial states, difficulty of adding new states, and numerical instability of the optimization procedure if proper regularization is not applied~\cite{Koch:1993,Kolda:2009,
Espig:2012,Hackbusch:2012,Goto:2013}.

We present an alternative technique that selects the important configurations in a black-box manner and is naturally regularized to prevent numerical instability.  It is built through a combination of imaginary-time evolution and compressed sensing.  Imaginary-time evolution can be concisely described as follows.  Given a quantum system with a time-independent Hamiltonian $H$ and associated eigenvectors $\left\{ \ket{\chi^i} \right\}$, any state of the system $\ket{\Omega}$ may be expressed in terms of those eigenvectors as
\begin{align}
 \ket{\Omega} = \sum_i c_i \ket{\chi^i}
\end{align}
and the the evolution of the system for imaginary-time $\tau$ is given as
\begin{align}
 G\ket{\Omega} = e^{-H \tau} \ket{\Omega} = \sum_i c_i e^{-E_i \tau} \ket{\chi^i}
\end{align}
where $E_0 < E_1 \leq E_2... \leq E_{N-1}$ are the eigenenergies associated with $\ket{\chi_i}$.  By evolving and renormalizing, eventually one is left with only the eigenvector associated with the lowest eigenvalue, or ground state.  Excited states may be obtained with a number of approaches including spectral transformations (e.g. $H'=(H-\lambda)^2$~\cite{MacDonald:1934}), matrix deflation, or other techniques. However we will concern ourselves only with ground states in this work.  

Imaginary time evolution approaches may be broadly grouped into two classes.  The first class involves the explicit application of the imaginary-time propagator $G$ to the wavefunction.  This approach typically generates many configurations at every step, causing a rapid expansion in the size of the wavefunction.  As a result, these methods have almost exclusively been restricted to Monte Carlo sampling procedures which attempt to assuage this explosion by stochastically sampling or selecting the most important configurations~\cite{Lester:1994,Booth:2009}, however the recently developed imaginary time-evolving block decimation also belongs to this class, performing truncations after expansion along a virtual bond dimension~\cite{Vidal_2004,Schollw_ck_2005,Clark:2014,Haegeman_2014}.  

The second class of imaginary-time approaches follow the evolution dictated by the action of $G$ projected onto the manifold spanned by linear variations in the function at the previous time step, sometimes referred to as Galerkin or time-dependent variational methods including imaginary time MCTDH~\cite{Beck:2000,Nest:2005} and DMRG in some limits~\cite{Haegeman_2014}.  While computationally convenient, it is often unclear how projection onto the original linear subspace at every time can affect evolution with respect to the exact evolution.  In this work, we show that the first class of explicit evolution can be used on any ansatz without configuration explosions or stochastic sampling by utilizing a technique from the field of compressed sensing, namely orthogonal matching pursuit~\cite{Pati:1993,Tropp:2007}.

\begin{figure}
\includegraphics[width=8.0 cm]{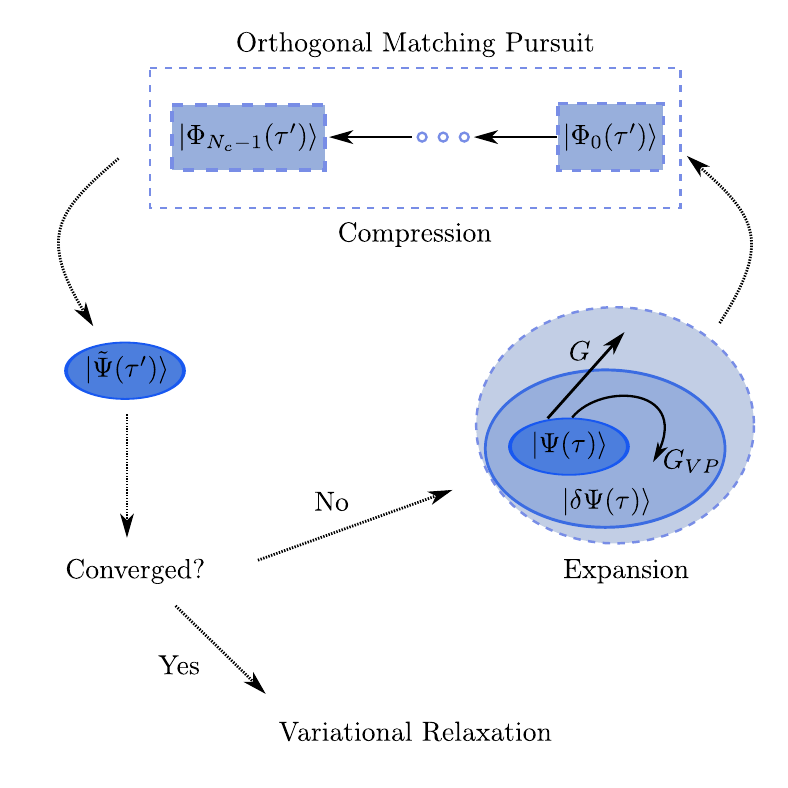}
\caption{A schematic diagram of the MAGIC approach.  At each iteration the wavefunction is expanded by means of the imaginary time propagator $G$, and subsequently compressed with orthogonal matching pursuit. The imaginary time propagator corresponding to projection into the manifold at $\ket{\Psi(\tau)}$, denoted $\ket{\delta \Psi(\tau)}$, typically prescribed by differential time dependent variational principles is given as $G_{VP}$ and depicted to emphasize that expansion with the operator $G$ can explore a greater part of Hilbert space. The compression is performed simultaneously with expansion in our implementation to prevent rapid growth of the wavefunction.  These steps are iterated until convergence at a specified maximum number of component functions, at which point an optional variational relaxation may be performed.}
\label{fig:MethodSketch}
\end{figure}
The algorithm we use is diagrammed in Fig \ref{fig:MethodSketch}, and proceeds iteratively as follows. The wavefunction at time $\tau=0$, $\ket{\Psi(\tau)}$, may be any trial wavefunction that is not orthogonal to the desired eigenstate.  We determine the wavefunction at time $\tau+d\tau=\tau'$ greedily, fitting one configuration $\ket{\Phi^i(\tau')}$ at a time by maximizing the functional
\begin{align}
\frac{ \left| \bra{\Phi^i(\tau')} G \ket{\Psi(\tau)} - \sum_{j < i} c_j(\tau') \braket{\Phi^i(\tau')}{\Phi^j(\tau')} \right|}{ \sqrt{ \braket{ \Phi^i(\tau')}{\Phi^i(\tau')}}}
\end{align}
with respect to the parameters that determine $\ket{\Phi^i(\tau')}$. Such that after $k$ iterations, the wavefunction is given by
\begin{align}
 \ket{\tilde \Psi(\tau)} = \sum_i^{k} c_i(\tau) \ket{\Phi^i(\tau)}
\end{align}
The coefficients in this expansion, $c_i(\tau')$ are solved for simultaneously after each iteration by orthogonal projection, which after simplification reduces to the following linear system for the coefficient vector $c$
\begin{align}
S c = v
\end{align}
where $S_{ij} = \braket{\Phi^i(\tau')}{\Phi^j(\tau')}$ and $v_{i} = \bra{\Phi^i(\tau')} G \ket{\Psi(\tau)}$.  Together, the fit and orthogonal projection step is equivalent to orthogonal matching pursuit~\cite{Pati:1993,Tropp:2007} applied to the signal generated by the imaginary time evolution of the state at each time step $G \ket{\Psi(\tau)}$. The expansion-compression procedure is advanced to the next imaginary time step either when some accuracy convergence criteria is met, or when some pre-set maximum number of components $N_c$ is reached, and the total simulation is terminated when the wavefunction converges between imaginary-time steps.  We provide additional details of the numerical procedure in the supporting information for interested readers.

Note that one is free to choose a convenient form for the propagator $G$. In this work we use the linearized propagator $G \approx (1 - d \tau H)$, which is both easy to implement and provably free of bias in the final result for finite single particle basis sets given some restrictions on $d\tau$~\cite{Trivedi:1990}.  

Orthogonal matching pursuit attempts to find the sparsest solution to the problem of state reconstruction~\cite{Tropp:2007,Needell:2009}, and thus is ideal for keeping the number of configurations minimal throughout the imaginary time evolution.  However, while the solution is sparsest in the limit of total reconstruction and naturally regularized against configuration collinearity, for very severe truncations of the wavefunction, the sparsifying conditions generate a solution which is not variationally optimal for the given number of configurations.  For this reason, we finish the computation with a total variational relaxation of the expectation value of the energy with respect to both coefficients and states.  This retains both the benefits of imaginary time evolution in avoiding local energetic minima and of variational optimality in the final solution.

\section{Application to chemical systems}
The method we have outlined may be readily applied to any quantum system, such as spins or oscillators, however as a first application we consider ground-state electronic wavefunctions of molecules. We will take the approach that is conventional to the field of quantum chemistry, and solve the problem in a basis of Gaussian-type functions~\cite{Helgaker:2002}.  After a basis has been selected, there is a standard procedure of expanding the linear state space by excitation known as configuration interaction (CI), which can eventually yield the numerically exact solution within a basis when the full state space has been covered.  This is referred to as full configuration interaction (FCI) and is the standard to which we compare.  Comparison to methods considering explicit correlation beyond that covered by a specific traditional Gaussian basis, such as explicitly correlated $f_{12}$ type wavefunctions, are not yet within the scope of this work. 

\begin{figure}
\includegraphics[width=8.0 cm]{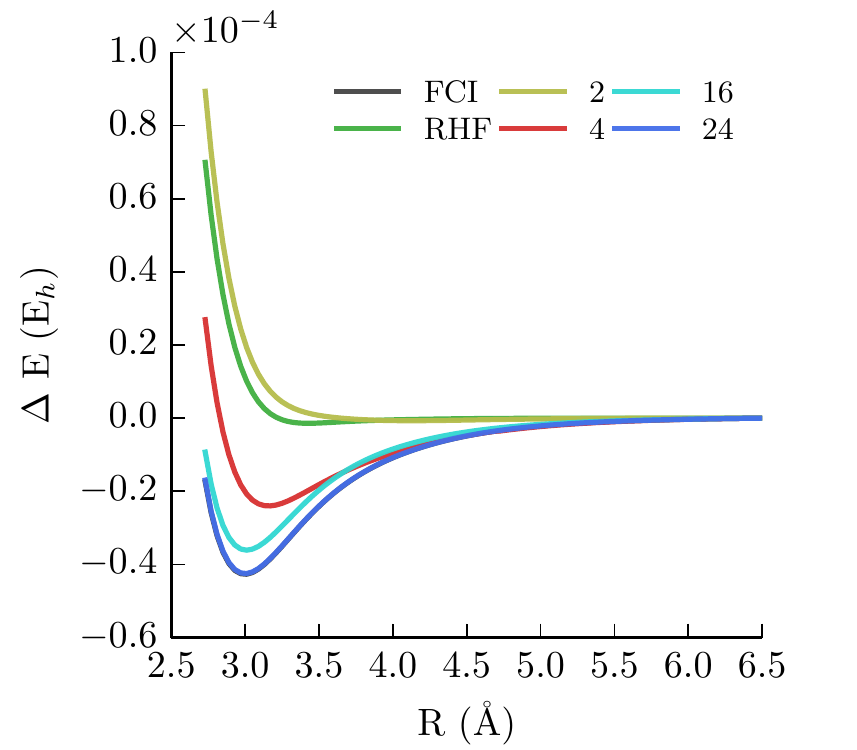}
\caption{The bond dissociation curve of the helium dimer in the aug-cc-pVDZ basis showing rapid and consistent convergence in the number of non-orthogonal determinants.  These represent the nuclear union curves constructed from a number of local determinants at each nuclear point given by the line label, and are sampled at a spacing of $0.04$ \AA.  The curves have been shifted by the tail values in order to allow comparison of the features for this sensitive bond, and the $24$ determinant curve is indistinguishable from the FCI solution in the graphic. At a point near the equilibrium geometry, $R=3.01$ \AA, the 24 determinant curve with the nuclear union configuration interaction technique recovers 99.9899\% of the basis set correlation energy, or equivalently 99.9999\% of the total energy.}
\label{fig:He2Curve}
\end{figure}

In the context of our approach, the indistinguishability of electrons necessitates handling of anti-symmetry.  The simplest way to include anti-symmetry into the wavefunction is by utilizing anti-symmetric component tensors $\ket{\Phi^i}$.  The most common anti-symmetric component function is the Slater determinant, such that we express the wavefunction as 
\begin{align}
 \ket{\Psi} = \sum_i^{N_{c}} c_i \ket{\Phi^i}
\end{align}
where $\ket{\Phi^i}$ are Slater determinants with no fixed relations between $\ket{\Phi^i}$ and $\ket{\Phi^j}$ for $i \neq j$. While this simple form lacks extensivity~\cite{Sundstrom:2014}, it is attractive for other reasons.  Namely the quality of description and rate of convergence in $N_c$ are invariant to invertible local transformations of the state (i.e. atomic orbitals vs. natural orbitals)~\cite{Hackbusch:2012}, and the mathematical machinery related to the use and extension of such a wavefunction is already well developed~\cite{HeadGordon:1998,Lowdin:1955,Amovilli:1997,Song:2009}


While the method we use for determinant selection is unique, non-orthogonal Slater determinants have been used successfully in valence bond theory~\cite{Hunt:1972,Goddard:1973} as well as more recent symmetry breaking and projection methods~\cite{Jime_nez_Hoyos_2013,Bytautas_2014}.  Unconstrained non-orthogonal Slater determinants have been utilized before, but in a purely variational context~\cite{Koch:1993,Goto:2013}. Using this machinery yields explicit gradients that we utilize in the optimization of determinants~\cite{Song:2009}.  The scaling of these constructions with current algorithms is $O({N_{c}}^2 \max(M^2, N_e^3))$~\cite{Sundstrom:2014} where ${N_{c}}$ is the number of determinants and $N_e$ is the number of electrons, however development of approximations in this area have received comparatively less attention with respect to orthogonal reference wavefunction methods and there may be ways to improve upon this scaling.

We introduce an additional enhancement for the study of chemical reactions that is greatly facilitated by the compactness of our expansions.  Namely, when considering a full reaction coordinate, such as that for a bond dissociation, we perform an additional linear variational calculation in the space of components (determinants) found locally at neighbouring nuclear configurations.  As a proof of principle, we include configurations from the entire curves in the following examples, but more economical truncations can be used as well.  We refer to this additional step, as the nuclear union configuration interaction method and describe it in more depth in the supplemental information.

As a first application, we consider He$_{2}$ in the aug-cc-pVDZ basis~\cite{Woon:1994}.  The helium dimer is unbound in the case of a single determinant with restricted Hartree Fock and is not held together by a covalent bond, but rather dispersive forces and dynamical correlation only.  In Fig. \ref{fig:He2Curve}, we consider the dissociation of this molecule under different numbers of non-orthogonal determinants.  Despite the sensitive nature of this bond, it is qualitatively captured with as few as $4$ local determinants and quantitatively captured with as few as $24$ determinants.  The dimension of the space of this molecule is on the order of $10^4$ when reduced by considerations of point group symmetry.  The NOMAGIC approach does not yet utilize any symmetry other than the spin symmetry enforced by the parameterization of the wavefunction.

\begin{figure}
\includegraphics[width=8.0 cm]{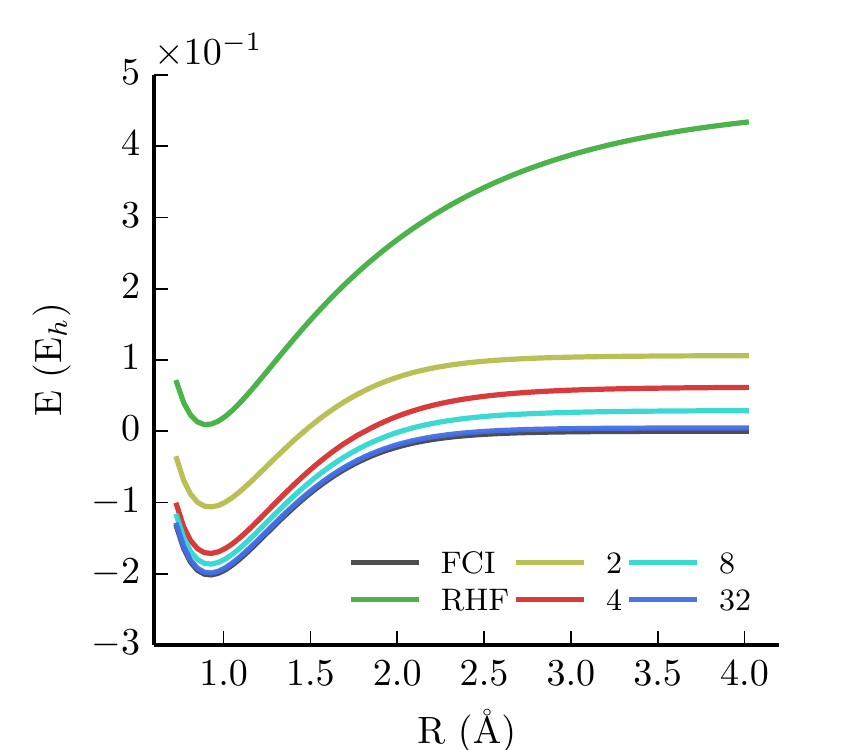}
\caption{The bond dissociation curve of hydrogen fluoride in the cc-pVDZ basis showing rapid convergence in the number of non-orthogonal determinants.  These are the nuclear union curves constructed from a number of local determinants at each nuclear point given by the line label, and are sampled at a spacing of $0.04$ \AA.  The $32$ determinant curve is nearly indistinguishable from the FCI curve in this graphic.  At a point near the equilibrium geometry, $R=0.93$ \AA, the 32 determinant curve with the nuclear union configuration interaction technique recovers 98.6\% of the basis set correlation energy, or equivalently 99.997\% of the total energy.}
\label{fig:HFCurve}
\end{figure}

As a second example, the dissociation of hydrogen fluoride in a cc-pVDZ basis~\cite{Dunning:1989} is studied.  The total configuration space for this molecule is on the order of $10^7$ and it involves a homolytic bond breaking of a covalent bond in the gas phase.  Considering the results in Fig. \ref{fig:HFCurve}, one can see that while restricted Hartree Fock (RHF) yields an unphysical dissociation solution, as few as $2$ determinants are sufficient to fix the solution in a qualitative sense.  Beyond this, the addition of more determinants represents a monotonically increasing degree of accuracy, with rapid convergence to a quantitative approximation by $32$ determinants.    

\begin{figure}
\includegraphics[width=8.0 cm]{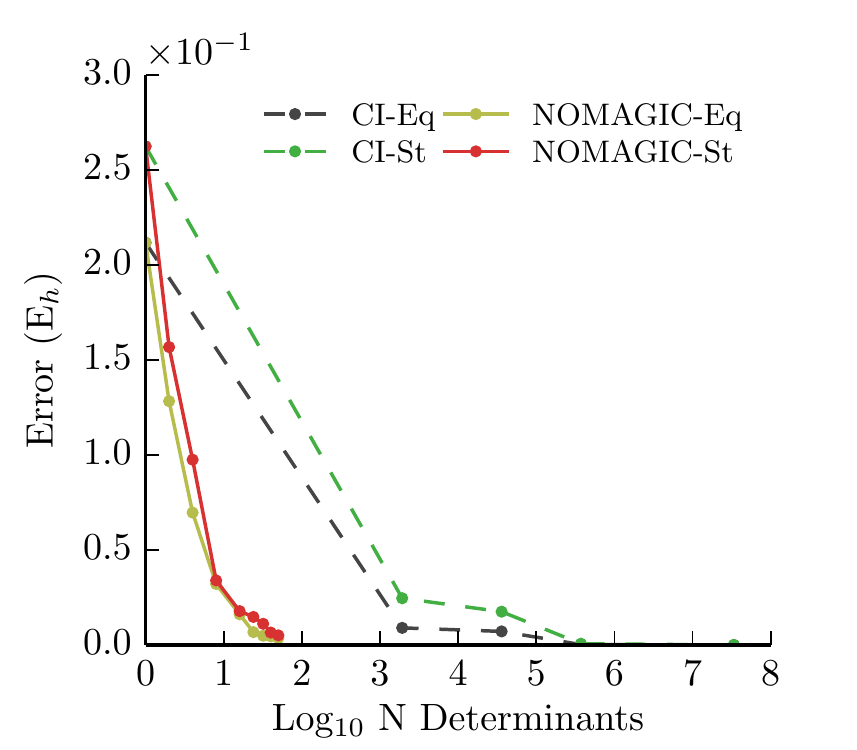}
\caption{A curve of the energetic error with respect to full CI for HF bond dissociation in the cc-pVDZ basis as a function of the log of the number of determinants included for both a near-equilibrium geometry ``Eq'' with $R=0.93$ \AA\ and a stretched geometry ``St'' with $R=1.73$ \AA.  The configuration interaction energies are generated by a standard excitation sequence from the Hartree-Fock solution, CIX(X=SD, SDT, SDTQ) followed by FCI.  The number of determinants used in the full CI expansion is approximately 34 million taking into account molecular point group symmetries, or 135 million without.  No symmetries other than spin are utilized in the NOMAGIC calculation.}
\label{fig:HFCompression}
\end{figure}

In Fig. \ref{fig:HFCompression} we select two points on the HF dissociation curve, and study the convergence of the energy as a function of the number of determinants in the NOMAGIC method and a traditional CI expansion with the canonical Hartree-Fock orbitals.  In particular, we study both a point near the equilibrium bond length ($R=0.93$ \AA) where traditional CI expansions perform relatively well and a more stretched geometry ($R=1.73$ \AA) where traditional CI expansions perform more poorly.  We see that in both cases, if one considers a fixed level of accuracy in the energy, the NOMAGIC method is considerably more compact.  For example, to achieve a level of accuracy superior to the CISDT expansion that uses $36021$ determinants, NOMAGIC requires only $24$ determinants at both geometries.  That is, for equivalent accuracy, the NOMAGIC wavefunction is roughly $1500$ times more compact in the space of Slater determinants.  By $50$ determinants out of a possible $10^7$ in the NOMAGIC wavefunction, we recover 98\% of the basis set correlation energy or equivalently 99.996\% of the total energy.

\section{Conclusions}
In this work, we introduced a general method to find compact representations of quantum eigenfunctions by using imaginary-time evolution and compression.  The method assumes no specific structure in the problem and can be applied to any quantum system with a variety of ansatze.  We demonstrated its success in some quantum chemical systems with a small number of non-orthogonal Slater determinants.  We believe that extensions to this method using ansatze that contain system specific physics have the potential to be even more compact and this is subject of current research.

\section{Acknowledgments}
We thank Prof. John Parkhill and Dr. Dmitrij Rappoport for their valuable comments on the manuscript.  J.M is supported by the Department of Energy Computational Science Graduate Fellowship under grant number DE-FG02-97ER25308.  A.A.G thanks the National Science Foundation for support under award CHE-1152291.

\section{Supplemental Information}

\subsection{Orthogonal Matching Pursuit}
In this section, we offer some additional details on the implementation of Orthogonal Matching Pursuit~\cite{Tropp:2007} with imaginary-time evolution in quantum systems.  Given a quantum state $\ket{\Psi(\tau')}$ that one wishes to reconstruct, orthogonal matching pursuit is a greedy decomposition algorithm that approximates the sparse problem of finding $\ket{\tilde \Psi(\tau')}$ such that 
\begin{align}
 \min_{\ket{\tilde \Psi(\tau')} } || \ket{\Psi(\tau')} - \ket{\tilde \Psi(\tau')}||^2_2 \notag \\
 \text{ \ \ subject to } ||\ket{\tilde \Psi(\tau')}||_0 < N
\end{align}
This is done by considering an overcomplete dictionary $\{\ket{\Phi^i(\tau')}\}$ that can express $\ket{\tilde \Psi}$ as
\begin{align}
\ket{\tilde \Psi(\tau')} = \sum_i c_i(\tau') \ket{\Phi^i(\tau')}
\end{align}
and at each stage selecting selecting the $\ket{\Phi^i(\tau')}$ which maximizes the overlap with the residual with respect to the target signal $\ket{\Psi(\tau')}$,
\begin{align}
 \max_{\ket{\Phi^i}(\tau')} \frac{|\braket{\Phi^i(\tau')}{\Psi(\tau')} - \sum_{j<i} c_j(\tau') \braket{\Phi^i(\tau')}{\Phi^j(\tau')}|}{\sqrt{\braket{\Phi^i(\tau')}{\Phi^i(\tau')}}}
\end{align}
In practice for quantum systems, the dictionary $\{ \ket{\Phi^i(\tau')} \}$ can be any overcomplete basis for the $N$-particle Hilbert space, and the location of the optimal $\ket{\Phi^i(\tau')}$ can be done with a few different methods such as discrete enumeration of all basis states, stochastic search, and direct non-linear optimization.  While discrete enumeration is commonly used in the orthogonal matching pursuit literature, the high dimensional nature of quantum systems does not readily allow it.  Among the remaining options, we find that direct direct non-linear optimization is superior to stochastic search methods for problems we considered.  Specifically, we utilized a quasi-Newton BFGS procedure with analytic gradients and inexact line search satisfying the strong Wolfe conditions.  

After selection of the optimal $\ket{\Phi^i(\tau')}$, the full set of coefficients $\{c_i(\tau')\}_{i=0}^{j}$ are re-determined by orthogonal projection of the selected basis functions on the signal $\ket{\Psi(\tau')}$.  This is equivalent to solving the linear equation 
\begin{align}
S c = v
\end{align}
for the coefficient vector $c$, where $S_{ij} = \braket{\Phi^i(\tau')}{\Phi^j(\tau')}$, $v_{i} = \braket{\Phi^i(\tau')}{\Psi(\tau')}$, and $c_i = c_i(\tau')$.  

Throughout this procedure, one also has a choice of how to represent the target signal $\ket{\Psi(\tau')}$.  In some cases, it is feasible to construct $\ket{\Psi(\tau')}$ explicitly from a previous time step and imaginary time propagator $G$, and doing so could potentially facilitate the optimization procedure by examining properties of the state.  However, exact expansion of the state $\ket{\Psi(\tau')}$ using $G$ can have many terms for even modestly sized quantum systems, negating the potential benefits of compressing the wavefunction.  In practice, we found that a much better approach is to directly with $G\ket{\Psi(\tau)}$ without first expanding the wavefunction explicitly. When using the linearized propagator $G\approx (1 - d\tau H)$, this means that Hamiltonian and overlap matrix elements and their derivatives are sufficient for the implementation of the procedure.

In principle, at any time step, one may continue to add elements $\ket{\Phi^i(\tau')}$ until an arbitrary convergence tolerance is reached, i.e. $|| \ket{\Psi(\tau')} - \ket{\tilde \Psi(\tau')} ||_2 < \epsilon$ for some $\epsilon > 0$.  However, as only the final state in the large $\tau$ limit is of interest, and any state that is not completely orthogonal to this state will eventually converge to it, some errors in intermediate steps are permissible.  Thus a more economical approach, is to terminate the addition of states $\ket{\Phi^i}$ at intermediate time steps according to some proxy, such as sufficient decrease in the energy $\tilde E(\tau') = \bra{\tilde \Psi(\tau')} H \ket{\tilde \Psi(\tau')}$ from the previous time step.

\subsection{Electronic Wavefunction Parameterization}
Here we detail the electronic wavefunction parametrization used in this work, as well as the expressions used for the implementation of orthogonal matching pursuit and variational relaxation in electronic systems.  

In quantum chemistry, frequently one first chooses a suitable single particle spin-orbital basis for the description of the electrons, which we denote $\{\ket{\phi_i}\}$.  This basis typically consists of atom-centered contracted Gaussian type functions with a spin function, and are in general non-orthogonal such that they have an overlap matrix defined by
\begin{align}
S_{ij} = \braket{\phi_i}{\phi_j}
\end{align}
Linear combinations of these atomic orbitals are used to form molecular orbital functions
\begin{align}
\ket{\chi_m} = \sum_i c_m^i \ket{\phi_i}
\end{align}
which have an inner product
\begin{align}
\braket{\chi_m}{\chi_n} = \sum_{i,j} c_m^{i*} c_n^i \braket{\phi_i}{\phi_j} = \sum_{i} c_m^{i*} c_n^j S_{ij}
\end{align}
In our implementation, the $N-$electron component wavefunctions may be formed from the anti-symmetrized $N-$fold product of molecular orbital functions, also known as Slater determinants.
\begin{align}
\ket{\Phi^k} = \mathcal{A} \left(\ket{\chi^k_0} \ket{\chi^k_1} ... \ket{\chi^k_{N-1}} \right) 
\end{align}
where $\mathcal{A}$ is the anti-symmetrizing operator.  A convenient computational representation  of an anti-symmetric component function $\ket{\Phi^k}$ is given by the coefficient matrix
\begin{align}
T^K = \left(c_0^K | c_1^K | ... | c_{N-1}^K \right)
\end{align} 
which denotes an $M\times N$ matrix whose $m$'th column are the coefficients defining the $m$'th molecular orbital $\ket{\chi^k_m}$.  This yields a convenient construction for the overlap between two component functions
\begin{align}
\braket{\Phi^K}{\Phi^L} = M_{KL} = \det \left( V_{KL} \right) = \det \left( T^{K\dagger} S T^L \right)
\end{align}
One quantity of convenience is the so-called transition density matrix defined between determinants $K$ and $L$
\begin{align}
 P^{KL} = T^{K} \left( T^{L \dagger} S T^K \right)^{-1} T^{L\dagger}
\end{align}
Hamiltonian matrix elements may be written as
\begin{align}
 H_{KL} &= M_{KL} \left(\Tr \left[P^{KL} \hat h\right] + \frac{1}{2}\Tr \left[ P^{KL} G^{KL} \right] \right)
\end{align}
where $\hat h$ are the single electron integrals,
\begin{align}
 h_{\mu \nu} = \int d\sigma\ \phi_\mu^*(\sigma) \left(-\frac{\nabla_{r}^2}{2} - \sum_{i} \frac{Z_i}{|R_i - r|} \right)\phi_\nu(\sigma) \\
\end{align}
where $\sigma=(r,s)$ denotes electronic spatial and spin variables and the nuclear positions and charges are $R_i$ and $Z_i$.
$G^{KL}$ is given by
\begin{align}
 G^{KL}_{\mu \nu} &= \left(\sum_{\lambda \sigma} P^{KL}_{\lambda \sigma} (g_{\mu \nu \lambda \sigma} - g_{\mu \lambda \nu \sigma}) \right)
\end{align}
with the two electron integrals $g_{\mu \nu \lambda \sigma}$
\begin{align}
g_{\mu \nu \lambda \sigma} = \int d\sigma_1\ d\sigma_2\ \frac{ \phi_\mu^*(\sigma_1) \phi_\nu(\sigma_1)  \phi_\lambda^*(\sigma_2) \phi_\sigma(\sigma_2) }{|r_1 - r_2|}
\end{align}

From the description of orthogonal matching pursuit, we see that to utilize non-linear optimization of the component functions $\ket{\Phi^k}$ with analytic first derivatives, one needs the variations of $H_{KL}$ and $M_{KL}$ with respect to $T^K$.  Allowing variations $\delta T^K$, the required expressions in the non-orthogonal spin orbital basis are as follows:
\begin{align}
 \delta M_{KL} &= M_{KL} \Tr \left[ S T^L (V^{KL})^{-1} \delta T^{K\dagger} \right] \\
 \delta P^{KL} &= [1 - P^{KL} S] \delta T^K (V^{KL\dagger})^{-1} T^{L\dagger} \\
 \delta G^{KL}_{\mu \nu} &= \left(\sum_{\lambda \sigma} \delta P^{KL}_{\lambda \sigma} (g_{\mu \nu \lambda \sigma} - g_{\mu \lambda \nu \sigma}) \right) \\
  A_{KL} &= \Tr \left[P^{KL} G^{KL}\right] \\
  \delta A_{KL}
  &=  \Tr \left[ (1 - SP^{KL\dagger}) G^{KL\dagger} T^{L} (V^{KL})^{-1} \delta T^{K\dagger} \right]
\end{align}
One must take care in implementing this expression, as it is a special case of the adjugate relations that is only strictly valid when $V^{KL}$ is non-singular.  To use this expression in evaluating cases when $V^{KL}$ is singular, techniques developed elsewhere utilizing the singular value decomposition of $V^{KL}$ and exact interpolation can be used~\cite{Amovilli:1997}.
Note also that numerical simplifications are possible by explicitly considering spin $(\alpha, \beta)$ and noting that $T^K = T^{K\alpha} \oplus  T^{K\beta}$.  These reductions of the above equations are straightforward and we do not give them here.

\subsection{Nuclear Union Configuration Interaction}
In this section we give some of the details of the nuclear union configuration interaction method used to improve the description of reaction coordinates.  In the study of a set of related problems, such as set of electronic Hamiltonians differing only by the positions of the nuclei, one would like to describe each configuration with an equivalent amount of accuracy, to get the best relative features possible.  In multi-reference methods, this is often done by selecting the same active space at each configuration, and rotating the orbitals and coefficients at each geometry accordingly.  In the nuclear union configuration interaction method, we propose each reuse of the components(determinants) found locally at other geometries to give a totally identical variational space for all nuclear configurations.  As the wavefunctions produced by the NOMAGIC method are especially compact, this introduces little extra overhead to the method as a whole.  

Specifically, denote the component functions found at nuclear configuration $R'$ with corresponding Hamiltonian $H(R')$ as $\ket{\Phi^k_{R'}} = \ket{\Phi^i}$ where $i$ is now an index set variable that runs over all the component functions at all the geometries being considered.  This could be a whole reaction coordinate, or simply neighbouring points depending on computational restrictions or chemical/physical considerations.  At each nuclear configuration $R$ we find new coefficients $c_i(R)$ by solving
\begin{align}
H(R)C = SCE
\end{align}
for its ground state eigenvector, and we define
\begin{align}
H(R)_{ij} &= \bra{\Phi^i} H(R) \ket{\Phi^j} \\
S_{ij} &= \braket{\Phi^i}{\Phi^j}
\end{align} 
Note that the overlap matrix may become singular, as configurations from nearby geometries are often very similar.  This can be handled either through canonical orthogonalization~\cite{Helgaker:2002} or by removing redundant configurations before attempting the diagonalization procedure.  Moreover, one might expect that additional compression is possible in this space, and this is the subject of current research.


\bibliographystyle{apsrev4-1}
\bibliography{CompactWfns}

\end{document}